\documentclass[
    ,final            % use final for the camera ready runs
  ]
  {aipproc}

\layoutstyle{6x9}

\begin{document}

\title{Polar Disk Galaxies as new way to study galaxy formation: the case of
NGC4650A}

\classification{98.62.Ai, 98.62.Bj, 98.62.Ck, 98.62.Dm, 98.62.Mw}
\keywords      {galaxies: formation --- galaxies: kinematics and dynamics}

\author{E. Iodice}{
  address={INAF-Astronomical Observatory of Capodimonte, Naples, Italy}
}

\begin{abstract}

NGC4650A is a polar disk galaxy: this is a peculiar object composed by
a central spheroidal component, the host galaxy (HG), and an extended disk
made up by gas, stars and dust, which orbits nearly perpendicular to
the plane of the central galaxy. The existence of two decoupled
components of the angular momentum let this object the ideal
laboratory i) to study gravitational interactions and merging and ii)
to constrain the 3D shape of its dark matter halo.  In view of these
applications, I will present two ongoing projects which aim to
constrain i) the formation scenario for polar disks and ii) the dark
halo content and shape, through a detailed analysis of the observed
structure, metallicity and dynamics of NGC4650A.

\end{abstract}

\maketitle

%%%%%%%%%%%%%%%%%%%%%%%%%%%%%%%%%%%%%%%%%%%%
%% MAINMATTER
%%%%%%%%%%%%%%%%%%%%%%%%%%%%%%%%%%%%%%%%%%%%

\section{The puzzling structure of NGC4650A}

In the latest years, our group has given a big effort to study the
formation and evolution of Polar Ring Galaxies (PRGs). NGC4650A is one
of the best-investigated object among the class of PRGs, in the South
hemisphere. Its luminous components, inner spheroid and polar
structure, have been studied with optical and near-infrared (NIR)
photometry and spectroscopy, and in the radio, HI 21 cm line and
continuum. The polar structure in NGC~4650A has been shown to be a
disk, rather than a ring. Its stars and dust can be reliably traced
inward within the stellar spheroid to $\sim 1.2$ kpc radius from the
galaxy nucleus (\citet{Iod02}; \citet{Gal02}). Emission and
absorption line optical spectra are also consistent with an extended
stellar disk rather than a narrow ring (\citet{SwR03}): both rotation
curves show a linear inner gradient and a plateau, as expected for a
disk in differential rotation. Furthermore, the HI 21 cm observations
(\citet{Arn97}) show that the gas is 5 times more extended
than the luminous polar structure, with a position-velocity diagram
very similar to those observed for edge-on gaseous disks. The polar
disk is very massive, since the total HI mass in this component is
about $10^{10} M_{\odot} $, which added to the mass of stars is
comparable with the total mass in the central spheroid
(\citet{Iod02}). New high resolution spectroscopy in NIR, obtained with
FORS2@UT4, along the main axes of the central HG suggests that it
resembles a nearly-exponential oblate spheroid supported by rotation
(\citet{Iod06}). These new spectroscopic data set important
constraints, both on current models for the formation scenarios and on
the mass models proposed till now for NGC4650A, which were the
starting points for the two ongoing works presented in the following
sections. They aimed i) to test the formation of polar disks by cold
accretion of gas through a "cosmic filament", and ii) to constrain the
dark halo (DH) content and shape.

\section{Chemical abundances in the polar disk: implications for the cold 
accretion} 

\begin{figure}
  \includegraphics[height=.3\textheight]{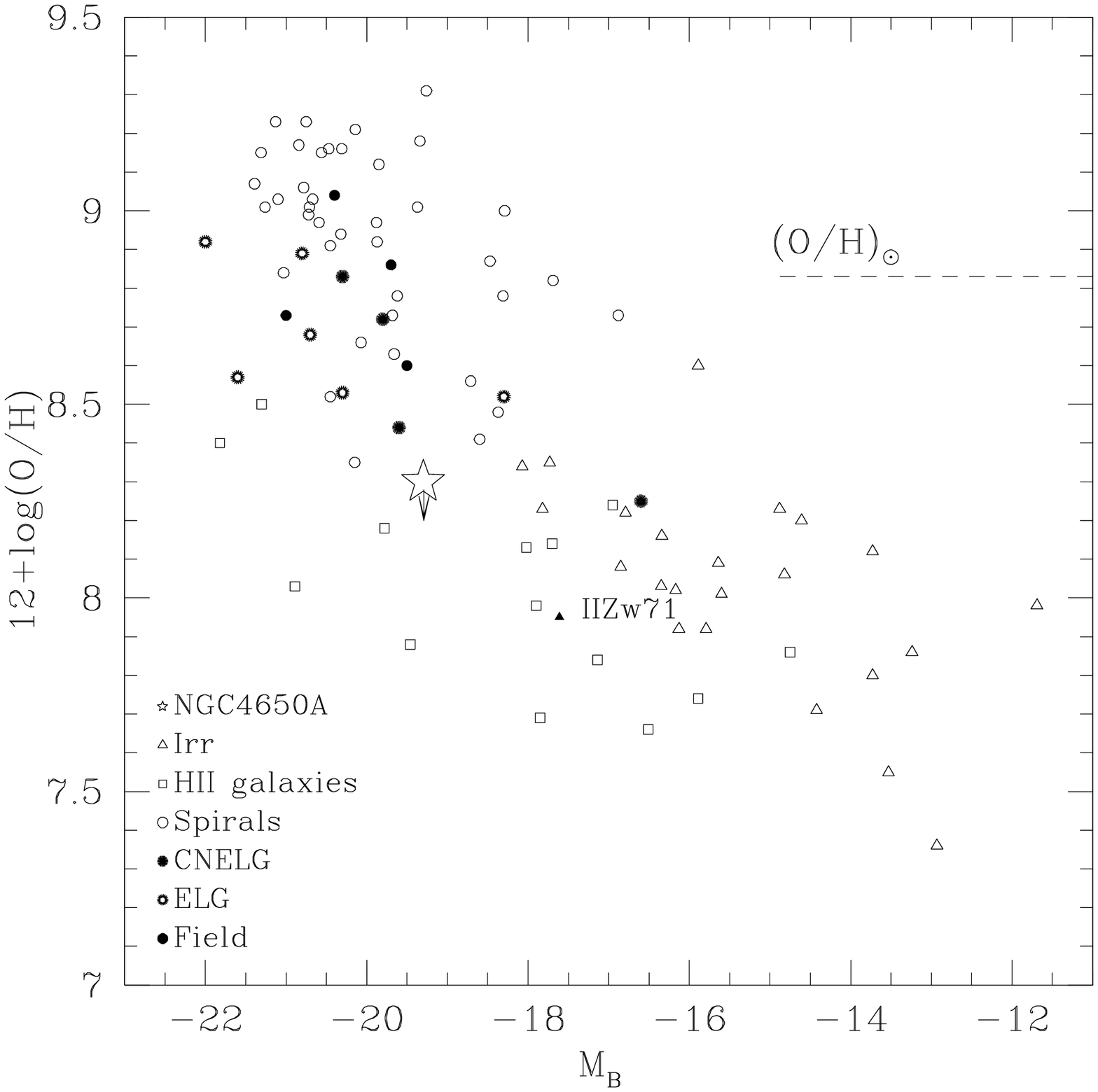} 
  \includegraphics[height=.3\textheight]{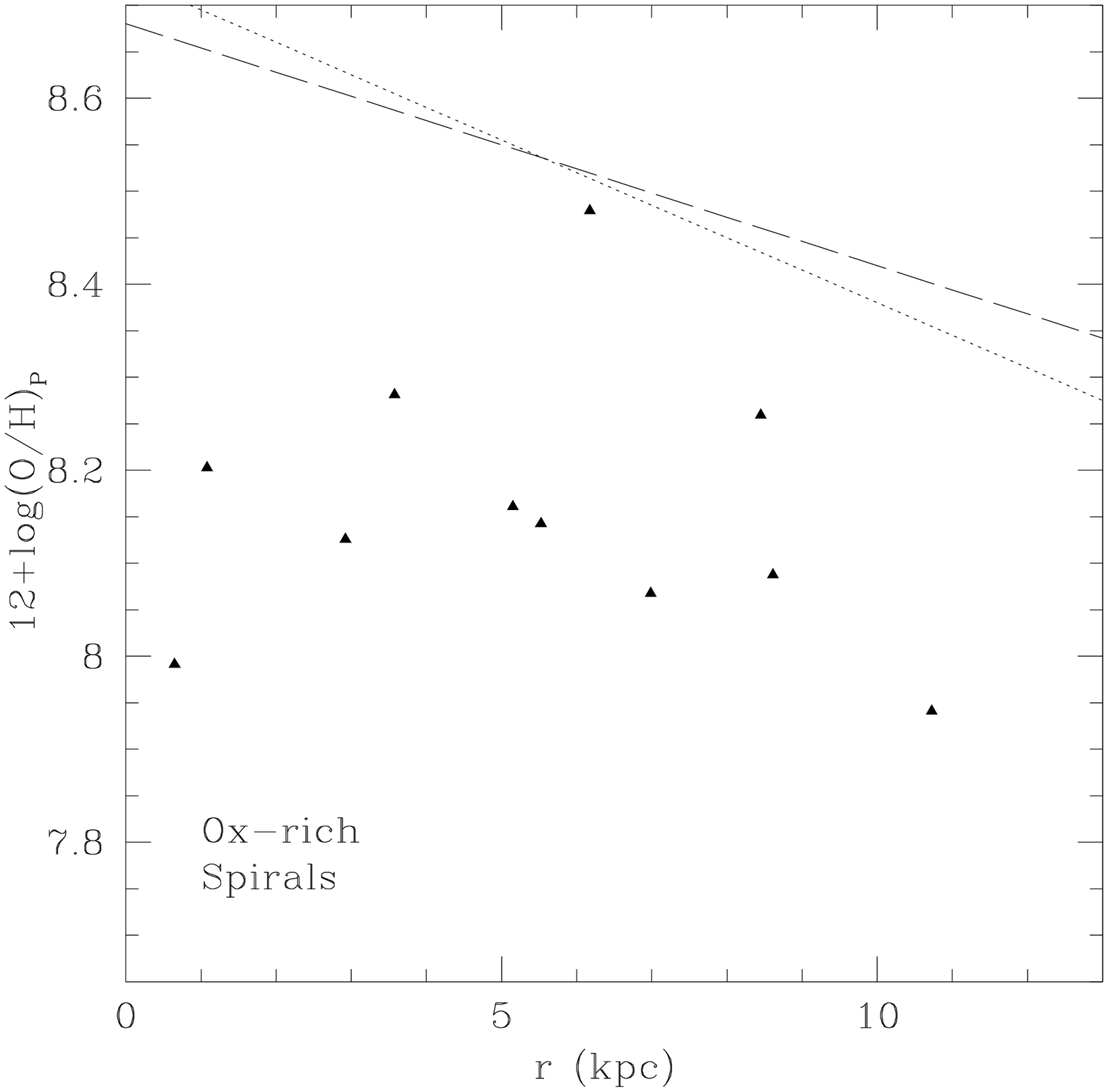} 
\caption{Left panel - Oxygen abundance vs absolute blue magnitude 
for samples of different morphological type of galaxies, including
Compact Narrow Emission Lines Galaxies (CNELG), Emission Lines
Galaxies (ELG), for NGC4650A (star) and the polar disk galaxy IIZw71
\citep{Per09}. The dashed line indicates the solar oxygen
abundance. The arrow indicates the shift of the value of the oxygen
abundance if we use the direct methods to evaluate it (see Spavone et
al. 2009 for details). Right panel - Oxygen abundance along the polar
disk derived with empirical methods (see Spavone et al. 2009 for
details). The superimposed lines are the linear best fit derived by
\citet{Pil06}; the dotted line represents the best fit to the abundance
of oxygen-rich spirals, while the long-dashed line is those related to
ordinary spirals.}\label{fig1}
\end{figure}

According to simulations of galaxy mergers (e.g. \citet{Bou05}),
polar merging of two disk galaxies fails to form a massive polar disk
around a spheroid with rotation velocities as large as observed along
the HG major axis (\citet{Iod06}); the tidal accretion scenario fails in
reproducing disk-like polar structures as extended as observed in
NGC4650A (\citet{Iod02}).  Very recently, a new formation mechanism has
been proposed within the scenario proposed for the build-up of high
redshift disk galaxies: a long-lived polar structure will form through
cold gas accretion along a filament, extended for $\sim 1$ Mpc, into
the virialized DH (\citet{Mac06}; \citet{Bro08}). In this
formation scenario, there is no limits to the mass of the accreted
material, thus a very massive polar disk may develop either around a
stellar disk or a spheroid: the morphology and kinematics of one
simulated object are quite similar to those observed for NGC4650A. In
order to test the cold accretion scenario for this object,
\citet{Spav09} (see also Spavone et al. in this book) have studied
the abundance ratios and metallicities of the HII regions associated
to the polar disk in NGC4650A, by using new deep longslit spectroscopy
obtained with FORS2@ESO-VLT. Main results show {\it i)} that NGC4650A
has metallicity lower than spiral galaxy disks of the same total
luminosity (see Fig.\ref{fig1} left panel), where $Z= 0.35 Z_{\odot}$,
which is consistent with values predicted for the formation of disks
by cold accretion processes; {\it ii)} the absence of any metallicity
gradient along the polar disk (see Fig.\ref{fig1} right panel), which
suggests that the metal enrichment is not influenced by the stellar
evolution of the older central spheroid and, thus, the disk was formed
later by the infall of metal-poor gas from outside which is still
forming the disk. 
%The latter is also a characteristic found in LSB
%galaxies (de Block \& van der Hulst 1998) and in some other PRGs
%(Broch et al. 2009).

\section{Dynamical model for NGC~4650A: solving the enigma
of the flattening of its Dark Halo}

\begin{figure}
  \includegraphics[height=.25\textheight]{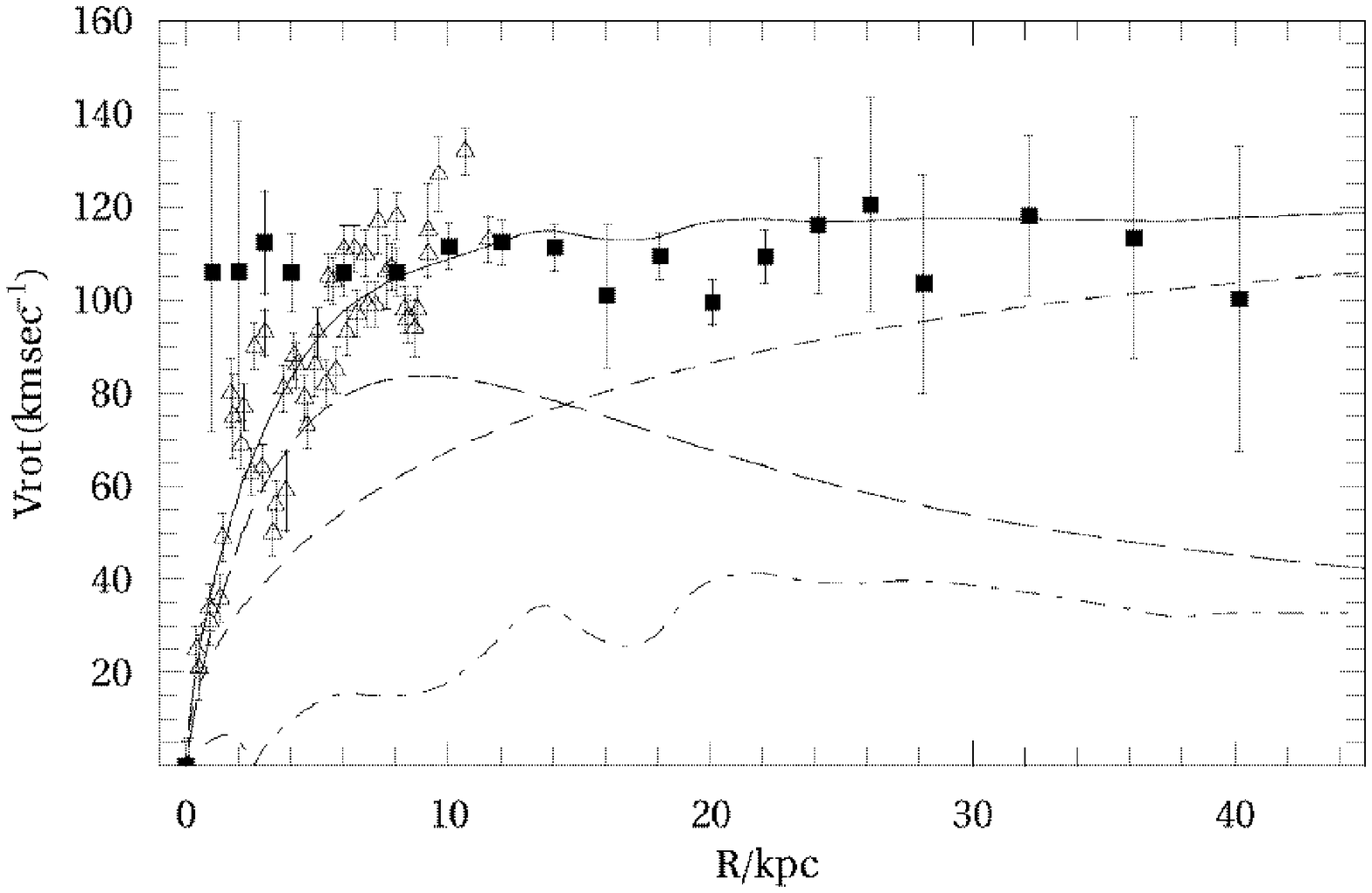} 
  \includegraphics[height=.25\textheight]{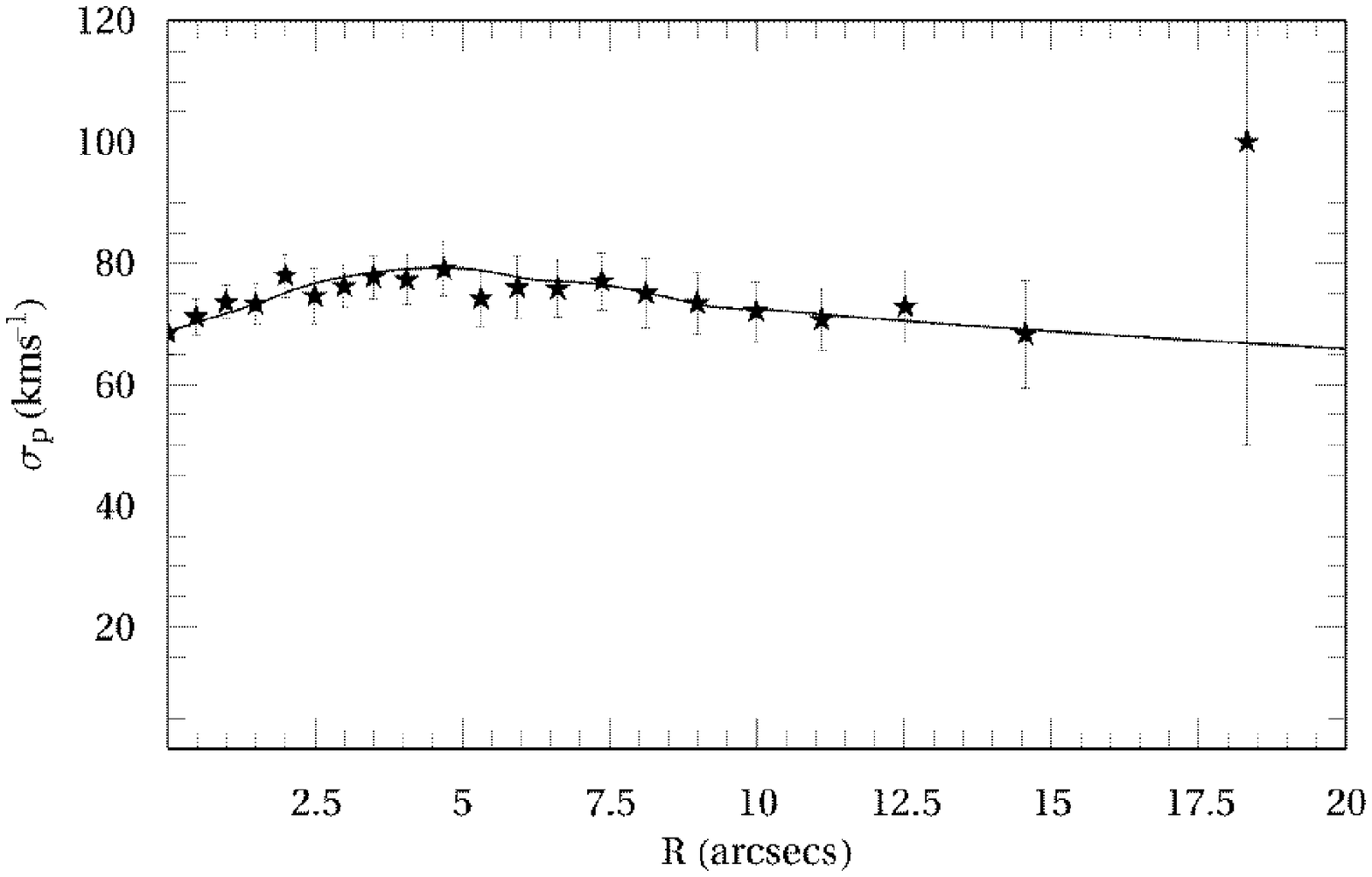} 
\caption{Left panel - Points are RC of stars (open triangles, by 
(\citet{SwR03}) and HI gas (filled squares, see \citet{Iod08}
for details); the continuous line is the best fit to the observed
velocities which accounts for an exponential disk (long-dash line), HI
gas density (short dash - long dash line) and a NFW DH (short-dash
line). Right panel - Points are the observed velocity dispersion along
the HG major axis (see \citet{Iod06}) and the continuous line is
the best fit obtained with a Jeans analysis. }\label{fig2}
\end{figure}

The existence of two orthogonal components of the angular momentum let
the PRGs the ideal laboratory to put limits on the 3-D shape of the DH.
The question of the DH shape is important because cosmological
simulations (\citet{NFW97}) predict the distribution of
the DH shapes and the universal radial dependence (NFW). NFW density
profiles are shown to describe the DH in many morphological types of
galaxies, but divergences have been found (\citet{dBl03};
\citet{Nap05}). In view of these applications, if one derives
the most likely DH flattening distribution from the PRGs dynamics,
tests can be performed on the likelihood of the different cosmological
models. The latest dynamical models developed for NGC4650A were by
\citet{Sack94}, that found a flattened (E6-E7) DH, whose axes
are aligned with those of the central galaxy, and by \citet{CA96} that
derived a best fit with a DH flattened in the plane of the polar disk
itself. The biggest uncertainties in the mass models proposed until
1996 are due to the low resolution of the HI data and to the
difficulty in measuring the velocity dispersion $\sigma$ along the
spheroid major axis. The new high resolution CaT spectra
(\citet{Iod06}) are better tracers of the kinematics for the
NGC4650A spheroid than was available before: they allow us to measure
a flat $\sigma$ profile along the spheroid major axis,
while previous measurements were too scattered to reliably
establish any trend with radius. The new $\sigma$ profiles show that
both the linear decreasing fit proposed by \citet{Sack94} and the
exponential empirical law proposed by \citet{CA96} do not reproduce
the observed trend with radius. Previous conclusions that the same
authors drew were based on data that are no longer valid, thus we aim
to develop a new dynamical model for NGC4650A (Iodice, Napolitano,
Arnaboldi et al. in preparation) by using the new spectroscopy
available for the HG (\citet{Iod06}) and rotation curves of stars
and gas in the polar disk (\citet{Arn97}; \citet{SwR03};
\citet{Iod08}).  As a first step we fitted the kinematics of the two
components (HG and polar disk) separately: 1) we have assumed a
centrifugal equilibrium along the polar disk, $V^2_c(R) = GM(R)/R$,
where the circular velocity $V^2_c = V^2_{star} + V^2_{gas} +
V^2_{DH}$; 2) we have performed an axisymmetric Jeans analysis on the
equatorial plane of the spheroid. The best fit for the polar disk (see
Fig.\ref{fig2} left panel) was obtained with $M/L=0.25$ (K band) by
adding to an exponential disk a DH with a NFW radial profile, with a
scale radius $r^{PD}_s = 54$ kpc and a DH mass of about 5.9 $10^{10}
M_{\odot}$ inside a radius of 25 kpc. To fit the rotation velocity and
velocity dispersion along the spheroid equatorial plane, we assumed a
NFW DH, with virial mass and the concentrations fixed from the polar
axis, and isotropy of the velocity ellipsoid: the best fit was
obtained with a scale radius $r^{HG}_s = 18$ kpc for the DH (see
Fig.\ref{fig2} right panel). By such a simple analysis we have
obtained that the two scale radii are different $r^{HG}_s / r^{PD}_s =
0.3$ which suggests a DH flattened along the polar axis with a
flattening of the potential $(c/a)_{\phi} \sim 0.7$ (E7). This
preliminary result, together with the DH mass, are consistent with the
DH content and flattening given by
\citet{CA96}. We are going to perform a more accurate dynamical model
to fit at the same time the kinematics along the equatorial and
meridian plane, by introducing the shape parameter in the DH profile.

%%%%%%%%%%%%%%%%%%%%%%%%%%%%%%%%%%%%%%%%%%%%
%% Sample figure:
%%
%% The option [height=...] scales the picture to the given height,
%% without it it would be printed at its nominal size
%%%%%%%%%%%%%%%%%%%%%%%%%%%%%%%%%%%%%%%%%%%%

%%%%%%%%%%%%%%%%%%%%%%%%%%%%%%%%%%%%%%%%%%%%

%%%%%%%%%%%%%%%%%%%%%%%%%%%%%%%%%%%%%%%%%%%%%%%%
%% BACKMATTER
%%%%%%%%%%%%%%%%%%%%%%%%%%%%%%%%%%%%%%%%%%%%%%%%

%\begin{theacknowledgments}
%\end{theacknowledgments}


%\end{document}


\bibliographystyle{aipproc}   % if natbib is available

\bibliography{001eiodice_biblio}

\IfFileExists{\jobname.bbl}{}
 {\typeout{}
  \typeout{******************************************}
  \typeout{** Please run "bibtex \jobname" to optain}
  \typeout{** the bibliography and then re-run LaTeX}
  \typeout{** twice to fix the references!}
  \typeout{******************************************}
  \typeout{}
 }

\end{document}